\documentclass[useAMS,usenatbib,A4]{mn2e}

\usepackage{graphicx}  
\usepackage{ amssymb }
\usepackage{journals}
\usepackage[dvips]{color}
\newcommand{\teff}{$T_{\rm eff}$} 
\newcommand{\logg}{$\log g$} 
\newcommand{\kms}{km s$^{-1}$}
\newcommand{\vt}{$\xi_t$} 
\newcommand{\fei}{Fe\,{\sc i}}
\newcommand{\feii}{Fe\,{\sc ii}}

\newcommand{\oi}{O\,{\sc i}}

\newcommand{\sii}{Si\,{\sc i}}
\newcommand{\cai}{Ca\,{\sc i}}

\newcommand{\tii}{Ti\,{\sc i}}

\newcommand{\nii}{Ni\,{\sc i}}
\newcommand{\cui}{Cu\,{\sc i}}

\newcommand{\baii}{Ba\,{\sc ii}}
\newcommand{\laii}{La\,{\sc ii}}

\newcommand{\euii}{Eu\,{\sc ii}}
\newcommand{\cri}{Cr\,{\sc i}}

\newcommand\simgt{\lower.3ex\hbox{\gtsima}}

\newcommand\msun{M$_{\odot}$}

\def\msun{$M_{\odot}$}

\usepackage{placeins}
\usepackage[modulo]{lineno}
\usepackage{hyperref}

\title[GRACES observations of young stars]{GRACES observations of young
[$\alpha$/Fe]-rich stars\thanks{Based on observations obtained with ESPaDOnS,
located at the Canada-France-Hawaii Telescope (CFHT). CFHT is operated by the
National Research Council of Canada, the Institut National des Sciences de
l'Univers of the Centre National de la Recherche Scientique of France, and the
University of Hawai'i. ESPaDOnS is a collaborative project funded by France
(CNRS, MENESR, OMP, LATT), Canada (NSERC), CFHT and ESA. ESPaDOnS was remotely
controlled from the Gemini Observatory, which is operated by the Association of
Universities for Research in Astronomy, Inc., under a cooperative agreement
with the NSF on behalf of the Gemini partnership: the National Science
Foundation (United States), the National Research Council (Canada), CONICYT
(Chile), the Australian Research Council (Australia), Minist\'{e}rio da
Ci\^{e}ncia, Tecnologia e Inova\c{c}\~{a}o (Brazil) and Ministerio de Ciencia,
Tecnolog\'{i}a e Innovaci\'{o}n Productiva (Argentina).}} 
\author[Yong et al.]
{David Yong,$^{1}$\thanks{E-mail: david.yong@anu.edu.au}
Luca Casagrande,$^{1}$ 
Kim A.\ Venn,$^2$ 
Andr{\' e}-Nicolas Chen{\' e},$^3$ \newauthor 
Jared Keown,$^{2}$
Lison Malo,$^{4}$ 
Eder Martioli,$^{5}$ 
Alan Alves-Brito,$^{6}$ 
Martin Asplund,$^{1}$\newauthor 
Aaron Dotter,$^{1}$
Sarah L.\ Martell,$^{7}$ 
Jorge Mel{\' e}ndez$^{8}$ and 
Katharine J.\ Schlesinger$^{1}$. \\
\\ 
$^{1}$Research School of Astronomy and Astrophysics, Australian
National University, Canberra, ACT 2611, Australia\\ 
$^{2}$Department of Physics and Astronomy, University of Victoria, Victoria, BC
V8W 3P2, Canada\\
$^{3}$Gemini Observatory, Northern Operations Centre, 670 North A'ohoku Place,
Hilo, HI, 96720, USA\\ 
$^{4}$Canada-France-Hawaii Telescope Corporation, 65-1238 Mamalahoa Highway,
Kamuela, HI 96743, USA\\ 
$^{5}$Laborat\'orio Nacional de Astrof\'{\i}sica (LNA/MCTI), Rua Estados
Unidos, 154, Itajub\'a, MG, Brazil\\ 
$^{6}$Instituto de Fisica, Universidade Federal do Rio Grande do Sul, Av. Bento
Goncalves 9500, Porto Alegre, RS, Brazil\\ 
$^{7}$School of Physics, University of New South Wales,
Sydney, NSW 2052, Australia\\ 
$^{8}$Departamento de Astronomia do IAG/USP, Universidade de S\~ao Paulo, Rua
do Mat\~ao 1226, Cidade Universit\'aria,\\ 05508-900 S\~ao Paulo, SP, Brazil\\ 
}
\begin{document}


\pagerange{\pageref{firstpage}$-$\pageref{lastpage}} \pubyear{2015}

\maketitle

\label{firstpage}

\begin{abstract}

We measure chemical abundance ratios and radial velocities in four massive
(i.e., young) [$\alpha$/Fe]-rich red giant stars using high-resolution high-S/N
spectra from ESPaDOnS fed by Gemini-GRACES. Our differential analysis ensures
that our chemical abundances are on the same scale as the
\citet{Alves-Brito:2010aa} study of bulge, thin and thick disk red giants. We
confirm that the program stars have enhanced [$\alpha$/Fe] ratios and are
slightly metal poor. Aside from lithium enrichment in one object, the program
stars exhibit no chemical abundance anomalies when compared to giant stars of
similar metallicity throughout the Galaxy.  This includes the elements Li, O,
Si, Ca, Ti, Cr, Ni, Cu, Ba, La, and Eu. Therefore, there are no obvious
chemical signatures that can help to reveal the origin of these unusual stars.
While our new observations show that only one star (not the Li-rich object)
exhibits a radial velocity variation, simulations indicate that we cannot
exclude the possibility that all four could be binaries. In addition, we find
that two (possibly three) stars show evidence for an infrared excess,
indicative of a debris disk. This is consistent with these young
[$\alpha$/Fe]-rich stars being evolved blue stragglers, suggesting their
apparent young age is a consequence of a merger or mass transfer. We would
expect a binary fraction of $\sim$50\% or greater for the entire sample of
these stars, but the signs of the circumbinary disk may have been lost since
these features can have short timescales. Radial velocity monitoring is needed
to confirm the blue straggler origin. 

\end{abstract}


\section{INTRODUCTION}

The atmospheres of low-mass stars retain, to a large extent, detailed
information on the chemical composition of the interstellar medium at the time
and place of their birth. The chemical abundance ratio [$\alpha$/Fe] has long
served as a key indicator of the relative contributions of different types of
stars and thus the degree of chemical enrichment
(e.g., \citealt{Tinsley:1979aa,Matteucci:1986aa,Venn:2004aa}). Massive stars
with short lifetimes that die as core collapse supernovae (SNe II) produce
$\alpha$-elements and modest amounts of Fe whereas longer lived thermonuclear
supernovae (SNe Ia) dominate the production of Fe-peak elements. Enhanced
[$\alpha$/Fe] ratios therefore indicate that the stars are relatively old such
that the gas from which they formed included SNe II contributions, but not
those from SNe Ia.  Indeed, stars with high [$\alpha$/Fe] ratios are generally
older than $\sim$8 Gyr (e.g., \citealt{Fuhrmann:2011aa,Bensby:2014aa}). 

Asteroseismology from the CoRoT \citep{Baglin:2006aa} and {\it Kepler}
\citep{Gilliland:2010aa} satellite missions have enabled accurate measurements
of stellar masses and radii based on standard seismic scaling relations for
stars with solar-like oscillations
\citep{Ulrich:1986aa,Brown:1991ab,Chaplin:2013aa}. Those mass determinations
greatly help to derive more robust age estimates (e.g.,
\citealt{Chaplin:2014aa,Lebreton:2014aa,Silva-Aguirre:2013aa,Silva-Aguirre:2015aa}).
For red giants in particular, their ages are determined to good approximation
by the time spent in the hydrogen burning phase, which is predominantly a
function of mass (e.g., \citealt{Miglio:2013aa,Casagrande:2016aa}). The
combination of chemical abundance measurements and asteroseismic information
has broadly confirmed that stars with higher overall metallicity, [Fe/H], and
solar [$\alpha$/Fe] ratios are young whereas stars with lower metallicity and
higher [$\alpha$/Fe] ratios are old. Additionally, \citet{Nissen:2015aa}
showed a strong correlation between [$\alpha$/Fe] and isochrone-based ages
among thin disk stars, with the oldest stars having the highest [$\alpha$/Fe]
ratios. 

A challenge to this general picture has emerged through the discovery of a
handful of stars with enhanced [$\alpha$/Fe] ratios and high masses that result
in young inferred ages \citep{Martig:2015aa,Chiappini:2015aa}. 
\citet{Martig:2015aa} identified a sample of 14 stars younger than 6 Gyr with
[$\alpha$/Fe] $\ge$ +0.13 based on high-resolution infrared spectra from APOGEE
(Apache Point Observatory Galactic Evolution Experiment;
\citealt{Majewski:2015aa}). These unusually high masses ($M$ $\gtrsim$ 1.4
\msun), and thus young ages, are robust to modifications to the standard
seismic scaling relations and to the assumption that the helium mass fractions
are low (i.e., primordial). While \citet{Epstein:2014aa} examined
potential issues in the scaling relations for metal-poor stars with [Fe/H] $<$
$-$1, \cite{Martig:2015aa} dismissed this possibility (see Section 7 in their
paper) and existing tests for red giant stars indicate that the masses are
likely accurate to better than $\sim$10\% \citep{Miglio:2013ab}. The spatial
distributions, radial velocities and guiding radii for the young $\alpha$-rich
stars are indistinguishable from the $\alpha$-rich population. Definitive
population membership, based on kinematics, is currently limited by the
proper-motion uncertainties. 

Possible explanations for the origin of these stars include ($i$) they are
evolved blue stragglers whose current masses lead to spurious age
determinations, ($ii$) they were formed during a recent gas accretion episode
in the Milky Way or ($iii$) they were born near the corotation radius near the
Galactic bar \citep{Martig:2015aa,Chiappini:2015aa}.  For the former
explanation, increasing the mass of a red giant from $\sim$1.0 to $\sim$1.4
\msun\ would lower the inferred age by about 5 Gyr \citep{Dotter:2008aa}, and
that amount of material is consistent with blue straggler formation scenarios
\citep{Sills:2009aa}. For the latter two explanations, the basic premise is
that the stars are genuinely young and that the gas from which they formed
remained relatively unprocessed reflecting mainly SNe II ejecta. Numerical
simulations with inhomogeneous chemical enrichment predict a small fraction of
young stars ($\sim$ 3 Gyr) with high [$\alpha$/Fe] ratios
\citep{Kobayashi:2011aa} and young metal-rich stars with high [$\alpha$/Fe]
ratios have been observed in the Galactic centre \citep{Cunha:2007aa}. 

The goal of this work is to confirm the [$\alpha$/Fe] ratios of four young
stars from \citet{Martig:2015aa}, identify any chemical signature that
may provide clues to the origin of these objects and measure radial velocities
to better understand the binary fraction. 

\section{SAMPLE SELECTION, OBSERVATIONS AND ANALYSIS}

The sample consists of four stars with [$\alpha$/Fe] $\ge$ +0.20 and ages $<$
4.0~Gyr from \citet{Martig:2015aa}, see Tables \ref{tab:obs} and
\ref{tab:param}. High resolution ($R$ = 67,500), high signal-to-noise ratio
(S/N $\simeq$ 150 - 300 per pixel near 6500\,\AA) optical spectra were taken
during initial science observations using the Gemini Remote Access to CFHT
ESPaDOnS \citep{Donati:2003aa} Spectrograph (GRACES; \citealt{Chene:2014aa}) in
June and July 2015 using the 1-fiber mode. Briefly, light from the Gemini
North telescope is fed to the CFHT ESPaDOnS spectrograph via two 270m-long
optical fibres with $\sim$8\% throughput (see \citealt{Chene:2014aa}). Details
of the observations are provided in Table \ref{tab:obs}. Data reduction was
performed using the OPERA pipeline (\citealt{Martioli:2012aa}, Malo et al.\ in
prep) and reduced spectra are available from the Gemini website ({\url
http://www.gemini.edu/sciops/instruments/july-2015-onsky-tests}). Subsequent
to those data being made publicly available, the OPERA pipeline was updated
by L.\ M. and the spectra were re-reduced\footnote{Our results and
conclusions are unchanged whether we use the publicly available or re-reduced
spectra.}. We used the unnormalised spectra without automatic correction of
the wavelength solution using telluric lines and co-added the individual
exposures for a given star.  Continuum normalisation was performed using
routines in {\sc iraf}\footnote{IRAF is distributed by the National Optical
Astronomy Observatories, which are operated by the Association of Universities
for Research in Astronomy, Inc., under cooperative agreement with the National
Science Foundation.}. 

\begin{table*}
 \centering
 \begin{minipage}{190mm}
  \caption{Details of the observations.} 
  \label{tab:obs} 
  \begin{tabular}{@{}lllllll@{}}
  \hline
        2MASS ID & 
        Filename & 
        OBSID & 
        Date/UT at end & 
        Exptime (s) &
        Airmass & 
        Instrument mode \\ 
\hline 
  \hline
J19081716+3924583  &  N20150604G0033.fits  &  GN-2015A-SV-171-9  &  2015-06-04/08:41:38  &  180  &  2.03  &  Spectroscopy, star only \\
J19081716+3924583  &  N20150604G0034.fits  &  GN-2015A-SV-171-9  &  2015-06-04/08:45:42  &  180  &  2.11  &  Spectroscopy, star only \\
J19081716+3924583  &  N20150604G0035.fits  &  GN-2015A-SV-171-9  &  2015-06-04/08:49:53  &  180  &  2.06  &  Spectroscopy, star only \\
  \\
J19093999+4913392  &  N20150721G0045.fits  &  GN-2015A-SV-171-11 &  2015-07-21/09:56:48  &  180  &  1.16  &  Spectroscopy, star only \\
J19093999+4913392  &  N20150721G0046.fits  &  GN-2015A-SV-171-11 &  2015-07-21/10:00:40  &  180  &  1.15  &  Spectroscopy, star only \\
J19093999+4913392  &  N20150721G0047.fits  &  GN-2015A-SV-171-11 &  2015-07-21/10:04:32  &  180  &  1.15  &  Spectroscopy, star only \\
  \\
J19083615+4641212  &  N20150721G0049.fits  &  GN-2015A-SV-171-13 &  2015-07-21/10:24:40  &  180  &  1.14  &  Spectroscopy, star only \\
J19083615+4641212  &  N20150721G0050.fits  &  GN-2015A-SV-171-13 &  2015-07-21/10:28:31  &  180  &  1.14  &  Spectroscopy, star only \\
J19083615+4641212  &  N20150721G0051.fits  &  GN-2015A-SV-171-13 &  2015-07-21/10:32:23  &  180  &  1.15  &  Spectroscopy, star only \\
   &   \\
J19101154+3915484  &  N20150721G0052.fits  &  GN-2015A-SV-171-15 &  2015-07-21/10:43:59  &  180  &  1.10  &  Spectroscopy, star only \\
J19101154+3915484  &  N20150721G0053.fits  &  GN-2015A-SV-171-15 &  2015-07-21/10:47:50  &  180  &  1.10  &  Spectroscopy, star only \\
J19101154+3915484  &  N20150721G0054.fits  &  GN-2015A-SV-171-15 &  2015-07-21/10:51:43  &  180  &  1.10  &  Spectroscopy, star only \\
  \hline
\end{tabular}
\end{minipage}
\end{table*}

The effective temperatures (\teff) for the program stars were determined using
the infrared flux method following \citet{Casagrande:2010aa,Casagrande:2014aa}.
The surface gravity (\logg) was determined from the masses and radii obtained
using the standard seismic scaling relations, and our \logg\ values are
essentially identical to those of \citet{Pinsonneault:2014aa}; the average
difference in \logg\ was 0.010 $\pm$ 0.007 ($\sigma$ = 0.013). Equivalent
widths (EWs) were measured by fitting Gaussian functions using routines in {\sc
iraf} and {\sc daospec} \citep{Stetson:2008aa}. The two sets of equivalent
width measurements were in excellent agreement ($<${\sc iraf} $-$ {\sc
daospec}$>$ = 0.8~m\AA; $\sigma$ = 1.6~m\AA) and were averaged (see Table
\ref{tab:ew}). (The minimum and maximum EWs used in the analysis were 7~m\AA\
and 125~m\AA, respectively.) Chemical abundances were obtained using the local
thermodynamic equilibrium (LTE) stellar line analysis program {\sc moog}
\citep{Sneden:1973aa,Sobeck:2011aa} and one-dimensional LTE model atmospheres
with [$\alpha$/Fe] = +0.4 from \citet{Castelli:2003aa}. The microturbulent
velocity (\vt) was estimated by forcing no trend between the abundance from
\fei\ lines and the reduced equivalent width. We required that the derived
metallicity be within 0.1 dex of the value adopted in the model atmosphere. The
final stellar parameters are presented in Table \ref{tab:param}. We estimate
that the internal uncertainties in \teff, \logg\ and \vt\ are 50~K, 0.05~cgs
and 0.2~\kms, respectively. Our stellar parameters are in good agreement with
the values published in \citet{Martig:2015aa}; the average differences in
\teff, \logg\ and [Fe/H] are 2 $\pm$ 66~K, $-$0.04 $\pm$ 0.05~cgs and $-$0.09
$\pm$ 0.02~dex, respectively. 

\begin{table*}
 \centering
 \begin{minipage}{190mm}
  \caption{Stellar parameters.} 
  \label{tab:param} 
  \begin{tabular}{@{}llcccrcc@{}}
  \hline
        KIC ID & 
        2MASS ID & 
        \teff & 
        \logg & 
        \vt & 
        [Fe/H] &
        Mass\footnote{These values are taken directly from
\citet{Martig:2015aa}. The masses are from the scaling relations.} & 
        Age$^a$ \\ 
        & 
        & 
        (K) & 
        (cgs) & 
        (\kms) & 
        (dex) & 
        (\msun) & 
        (Gyr) \\
\hline 
  \hline
4350501  & J19081716+3924583 & 4689 & 3.05 & 1.01 & $-$0.14  & 1.65 $\pm$ 0.20 & $<$3.0  \\
9821622  & J19083615+4641212 & 4895 & 2.71 & 1.18 & $-$0.40  & 1.71 $\pm$ 0.26 & $<$2.6  \\
11394905 & J19093999+4913392 & 4951 & 2.50 & 1.38 & $-$0.51  & 1.40 $\pm$ 0.18 & $<$4.0  \\
4143460  & J19101154+3914584 & 4711 & 2.50 & 1.26 & $-$0.39  & 1.58 $\pm$ 0.20 & $<$3.1  \\
\multicolumn{8}{c}{Comparison field giant} \\
HD 40409 & J05540606-6305230 & 4746 & 3.20 & 1.19 &   +0.22  &          \ldots & \ldots  \\
  \hline
\end{tabular}
\end{minipage}
\end{table*}

\begin{table*}
 \centering
 \begin{minipage}{180mm}
  \caption{Line list for the program stars\label{tab:ew}}
  \begin{tabular}{@{}cccrrrrrrrrrrrrrrrc@{}}
  \hline
Wavelength & 
Species\footnote{The digits to the left of the decimal point are the atomic
number. The digit to the right of the decimal point is the ionization state
(``0'' = neutral, ``1'' = singly ionised).} & 
L.E.P & 
$\log gf$ & 
KIC 4350501 &
KIC 9821622 & 
KIC 11394905 & 
KIC 4143460 & 
HD 40409 & 
Source\footnote{A = $\log gf$ values used in \citet{Yong:2005aa} where the
references include \citet{Ivans:2001aa}, \citet{Kurucz:1995aa},
\citet{Prochaska:2000ab}, \citet{Ramirez:2002aa}; 
B = \citet{Gratton:2003aa};
C = Oxford group including 
\citet{Blackwell:1979aa,Blackwell:1979ab,Blackwell:1980aa,Blackwell:1986aa,Blackwell:1995aa}; 
D = \citet{Fuhr:2006aa}, using line component patterns for 
     hfs/IS from \citet{Kurucz:1995aa}; 
E = \citet{Fuhr:2006aa}, using hfs/IS from \citet{McWilliam:1998aa};
F = \citet{Lawler:2001aa}, using hfs from \citet{Ivans:2006aa};
G = \citet{Lawler:2001ab}, using hfs/IS from \citet{Ivans:2006aa}.  
\\
~
\\ 
This table is published in its entirety in the electronic edition of the paper.
A portion is shown here for guidance regarding its form and content.} \\ 
\AA & 
 & 
eV & 
 & 
m\AA & 
m\AA & 
m\AA & 
m\AA & 
m\AA & 
 \\ 
(1) & 
(2) &
(3) &
(4) & 
(5) & 
(6) & 
(7) & 
(8) & 
(9) & 
(10) \\ 
\hline
 6300.31 &    8.0 &   0.00 & $-$9.75 &   \multicolumn{5}{c}{Spectrum synthesis}             &           B \\
 7771.95 &    8.0 &   9.15 &    0.35 &     38.5 &     39.2 &   \ldots &     45.6 &     29.4 &           B \\
 7774.18 &    8.0 &   9.15 &    0.21 &     38.0 &     30.6 &     39.3 &     43.6 &     33.9 &           B \\
 5665.56 &   14.0 &   4.92 & $-$2.04 &     56.0 &     48.7 &     46.3 &     57.4 &     67.5 &           B \\
 5684.49 &   14.0 &   4.95 & $-$1.65 &     69.2 &   \ldots &   \ldots &   \ldots &   \ldots &           B \\
\hline
\end{tabular}
\end{minipage}
\end{table*}

Chemical abundances for other elements were obtained using the measured EWs,
final model atmospheres and {\sc moog}. For the 6300\,\AA\ [OI] line, Cu and
the neutron-capture elements, abundances were determined via spectrum synthesis
and $\chi^2$ minimisation. For the 6300\,\AA\ [OI] line, our line list
includes the Ni blend (which only contributes $<$20\% to the total equivalent
width). The abundances from the 6300\,\AA\ [OI] line are, on average, lower
than the abundances from the 777nm O triplet by $\sim$ 0.3 dex. That difference
is in good agreement with the non-local thermodynamic equilibrium (NLTE)
corrections by \citet{Amarsi:2015aa} (when assuming A(O)$_{\rm LTE}$ = 8.8,
\teff\ = 5000 K, \logg\ = 3.0 and [Fe/H] = $-$0.5; their grid does not yet
extend to lower \teff\ and lower \logg). For Cu, Ba, La and Eu, we included
isotopic shifts (IS) assuming solar abundances and hyperfine structure (hfs) in
the line lists. The chemical abundances are presented in Tables \ref{tab:abun}
and \ref{tab:abun2}. We adopted solar abundances from \citet{Asplund:2009aa}
and the uncertainties were determined following the approach in
\citet{Yong:2014aa}. The abundance uncertainties from errors in the stellar
parameters are provided in Table \ref{tab:err}. For the majority of lines,
we used damping constants from \citet{Barklem:2000aa} and
\citet{Barklem:2005ab}. For the remaining lines, we used the
\citet{Unsold:1955aa} approximation. 

A bright comparison giant star (HD 40409) studied by \citet{Alves-Brito:2010aa}
was also included in our analysis.  After comparing our abundance ratios with
those of \citet{Alves-Brito:2010aa} for the comparison star HD
40409\footnote{Stellar parameters for the comparison star were taken from
\citet{Alves-Brito:2010aa} as we were unable to apply the same methods as for
the program stars.}, we made minor offsets\footnote{The average offset was
$-$0.06 and the individual values were: \fei\ (+0.01), \feii\ ($-$0.15), \oi\
($-$0.17), \sii\ ($-$0.18), \cai\ (+0.11), and \tii\ (+0.04). For \feii,
\oi\ and \sii, the differences are non-negligible and likely due to differences
in the line selection and atomic data.} to place our Fe and $\alpha$-element
abundances onto their scale to aid our interpretation in the following
sections; \citeauthor{Alves-Brito:2010aa} did not report abundances for the
other elements. 

We re-analysed the stars using ATLAS9 model atmospheres generated by
\citet{Meszaros:2012aa}. When compared to the ATLAS9 models by
\citet{Kurucz:1993aa} and \citet{Castelli:2003aa}, the newer models include an
updated H$_2$O line list, a larger range of carbon and $\alpha$-element
abundances and solar abundances from \citet{Asplund:2009aa}. The average
difference in [X/Fe] ratios (\citealt{Meszaros:2012aa} (with [C/Fe] = 0 and
[$\alpha$/Fe] = +0.3) $-$ \citealt{Castelli:2003aa}) was only $-$0.03 $\pm$
0.01 dex. 

\begin{table}
 \centering
 \begin{minipage}{190mm}
  \caption{Chemical abundances for the program stars (\oi\ - \feii).}
  \label{tab:abun} 
  \begin{tabular}{@{}lcrcrc@{}}
  \hline
        Name & 
        A(X) & 
        N$_{\rm lines}$ & 
        s.e.$_{\log\epsilon}$ & 
        [X/Fe] & 
        $\sigma$[X/Fe] \\
\hline 
  \hline
     \multicolumn{6}{c}{\oi}   \\ 
KIC 4350501          &    9.07 & 3      &    0.05 &    0.53 &    0.16 \\ 
KIC 9821622          &    8.40 & 3      &    0.08 &    0.11 &    0.13 \\ 
KIC 11394905         &    8.24 & 2      &    0.10 &    0.07 &    0.15 \\ 
KIC 4143460          &    8.90 & 2      &    0.03 &    0.60 &    0.14 \\ 
HD 40409             &    9.01 & 2      &    0.14 &    0.10 &    0.17 \\ 
     \multicolumn{6}{c}{\sii}   \\ 
KIC 4350501          &    7.55 & 16     &    0.02 &    0.18 &    0.06 \\ 
KIC 9821622          &    7.17 & 15     &    0.02 &    0.06 &    0.07 \\ 
KIC 11394905         &    7.05 & 13     &    0.03 &    0.05 &    0.06 \\ 
KIC 4143460          &    7.30 & 12     &    0.03 &    0.18 &    0.06 \\ 
HD 40409             &    7.74 & 10     &    0.05 &    0.01 &    0.07 \\ 
     \multicolumn{6}{c}{\cai}   \\ 
KIC 4350501          &    6.23 & 9      &    0.04 &    0.04 &    0.07 \\ 
KIC 9821622          &    6.30 & 10     &    0.03 &    0.36 &    0.06 \\ 
KIC 11394905         &    6.09 & 9      &    0.03 &    0.27 &    0.06 \\ 
KIC 4143460          &    6.20 & 10     &    0.03 &    0.25 &    0.08 \\ 
HD 40409             &    6.47 & 3      &    0.03 & $-$0.09 &    0.10 \\ 
     \multicolumn{6}{c}{\tii}   \\ 
KIC 4350501          &    4.83 & 25     &    0.02 &    0.02 &    0.07 \\ 
KIC 9821622          &    4.95 & 33     &    0.02 &    0.40 &    0.06 \\ 
KIC 11394905         &    4.73 & 20     &    0.02 &    0.29 &    0.05 \\ 
KIC 4143460          &    4.75 & 23     &    0.02 &    0.19 &    0.07 \\ 
HD 40409             &    5.14 & 19     &    0.03 & $-$0.03 &    0.08 \\ 
     \multicolumn{6}{c}{\cri}   \\ 
KIC 4350501          &    5.39 & 8      &    0.07 & $-$0.10 &    0.08 \\ 
KIC 9821622          &    5.31 & 7      &    0.06 &    0.07 &    0.06 \\ 
KIC 11394905         &    5.17 & 7      &    0.06 &    0.05 &    0.06 \\ 
KIC 4143460          &    5.25 & 6      &    0.08 &    0.00 &    0.09 \\ 
HD 40409             &    5.77 & 5      &    0.09 & $-$0.09 &    0.10 \\ 
     \multicolumn{6}{c}{\fei}   \\ 
KIC 4350501          &    7.34 & 98     &    0.01 & $-$0.16 &    0.08 \\ 
KIC 9821622          &    7.12 & 98     &    0.01 & $-$0.38 &    0.07 \\ 
KIC 11394905         &    7.00 & 99     &    0.01 & $-$0.50 &    0.07 \\ 
KIC 4143460          &    7.11 & 87     &    0.01 & $-$0.39 &    0.08 \\ 
HD 40409             &    7.71 & 58     &    0.01 &    0.21 &    0.09 \\ 
     \multicolumn{6}{c}{\feii}   \\ 
KIC 4350501          &    7.50 & 10     &    0.04 &    0.00 &    0.11 \\ 
KIC 9821622          &    6.93 & 10     &    0.03 & $-$0.57 &    0.10 \\ 
KIC 11394905         &    6.81 & 8      &    0.03 & $-$0.69 &    0.11 \\ 
KIC 4143460          &    7.12 & 12     &    0.03 & $-$0.38 &    0.11 \\ 
HD 40409             &    7.80 & 6      &    0.04 &    0.30 &    0.12 \\ 
  \hline
\end{tabular}
\end{minipage}
\end{table}

\begin{table}
 \centering
 \begin{minipage}{190mm}
  \caption{Chemical abundances for the program stars (\nii\ - \euii).}
  \label{tab:abun2} 
  \begin{tabular}{@{}lcrcrc@{}}
  \hline
        Name & 
        A(X) & 
        N$_{\rm lines}$ & 
        s.e.$_{\log\epsilon}$ & 
        [X/Fe] & 
        $\sigma$[X/Fe] \\
\hline 
  \hline
     \multicolumn{6}{c}{\nii}   \\ 
KIC 4350501          &    6.14 & 23     &    0.02 &    0.06 &    0.03 \\  
KIC 9821622          &    5.86 & 19     &    0.01 &    0.04 &    0.04 \\  
KIC 11394905         &    5.70 & 19     &    0.01 & $-$0.01 &    0.04 \\  
KIC 4143460          &    5.86 & 23     &    0.02 &    0.03 &    0.03 \\  
HD 40409             &    6.47 & 18     &    0.02 &    0.03 &    0.04 \\  
     \multicolumn{6}{c}{\cui}   \\ 
KIC 4350501          &    4.08 & 2      &    0.09 &    0.03 &    0.12 \\  
KIC 9821622          &    3.71 & 2      &    0.12 & $-$0.09 &    0.13 \\  
KIC 11394905         &    3.57 & 2      &    0.09 & $-$0.11 &    0.15 \\  
KIC 4143460          &    3.71 & 2      &    0.07 & $-$0.08 &    0.12 \\  
HD 40409             &    4.38 & 2      &    0.06 & $-$0.03 &    0.11 \\  
     \multicolumn{6}{c}{\baii}   \\ 
KIC 4350501          &    2.06 & 3      &    0.12 &    0.02 &    0.13 \\  
KIC 9821622          &    1.83 & 3      &    0.11 &    0.04 &    0.16 \\  
KIC 11394905         &    1.80 & 3      &    0.08 &    0.14 &    0.15 \\  
KIC 4143460          &    1.91 & 3      &    0.14 &    0.12 &    0.18 \\  
HD 40409             &    2.42 & 3      &    0.13 &    0.02 &    0.15 \\  
     \multicolumn{6}{c}{\laii}   \\ 
KIC 4350501          &    1.01 & 2      &    0.06 &    0.05 &    0.11 \\  
KIC 9821622          &    0.88 & 2      &    0.10 &    0.18 &    0.11 \\  
KIC 11394905         &    0.82 & 2      &    0.12 &    0.24 &    0.14 \\  
KIC 4143460          &    0.94 & 2      &    0.04 &    0.23 &    0.12 \\  
HD 40409             &    1.57 & 2      &    0.03 &    0.25 &    0.12 \\  
     \multicolumn{6}{c}{\euii}   \\ 
KIC 4350501          &    0.74 & 1      &  \ldots &    0.36 &    0.16 \\  
KIC 9821622          &    0.66 & 1      &  \ldots &    0.54 &    0.16 \\  
KIC 11394905         &    0.47 & 1      &  \ldots &    0.46 &    0.17 \\  
KIC 4143460          &    0.64 & 1      &  \ldots &    0.51 &    0.16 \\  
HD 40409             &    0.73 & 1      &  \ldots & $-$0.01 &    0.16 \\  
  \hline
\end{tabular}
\end{minipage}
\end{table}

\begin{table}
 \centering
 \begin{minipage}{80mm}
  \caption{Abundance errors from uncertainties in atmospheric parameters.}
  \label{tab:err} 
  \begin{tabular}{@{}lrrrrc@{}}
  \hline
        Species & 
        $\Delta$\teff & 
        $\Delta$\logg & 
        $\Delta$\vt & 
        $\Delta$[m/H] & 
        Total\footnote{The total error is determined by adding in quadrature
the first four entries. 
\\
This table is published in its entirety in the electronic edition of the paper.
A portion is shown here for guidance regarding its form and content.} \\
        & 
        +50 K  &
        +0.05 cgs &
        +0.2 \kms &
        +0.2 dex \\ 
\hline 
  \hline
\multicolumn{6}{c}{KIC 4350501} \\ 
$\Delta$[\oi/Fe]    & $-$0.10 &    0.04 &    0.06 & $-$0.04 &    0.13 \\
$\Delta$[\sii/Fe]   & $-$0.03 &    0.01 &    0.04 & $-$0.00 &    0.05 \\
$\Delta$[\cai/Fe]   &    0.05 & $-$0.01 & $-$0.01 & $-$0.01 &    0.05 \\
$\Delta$[\tii/Fe]   &    0.06 & $-$0.01 &    0.02 & $-$0.01 &    0.06 \\
$\Delta$[\cri/Fe]   &    0.04 & $-$0.01 &    0.01 & $-$0.01 &    0.04 \\
  \hline
\end{tabular}
\end{minipage}
\end{table}

\section{RESULTS} 

We confirm that the program stars are slightly more metal poor than the Sun and
have enhanced [$\alpha$/Fe] ratios ($\alpha$ is the average of O, Si, Ca
and Ti). That is, our independent study using optical spectra from GRACES
largely confirms the results from the APOGEE pipeline analysis of the infrared
H-band spectrum \citep{Majewski:2015aa,Garcia-Perez:2015aa}. 

\begin{figure}
\centering
      \includegraphics[width=.99\hsize]{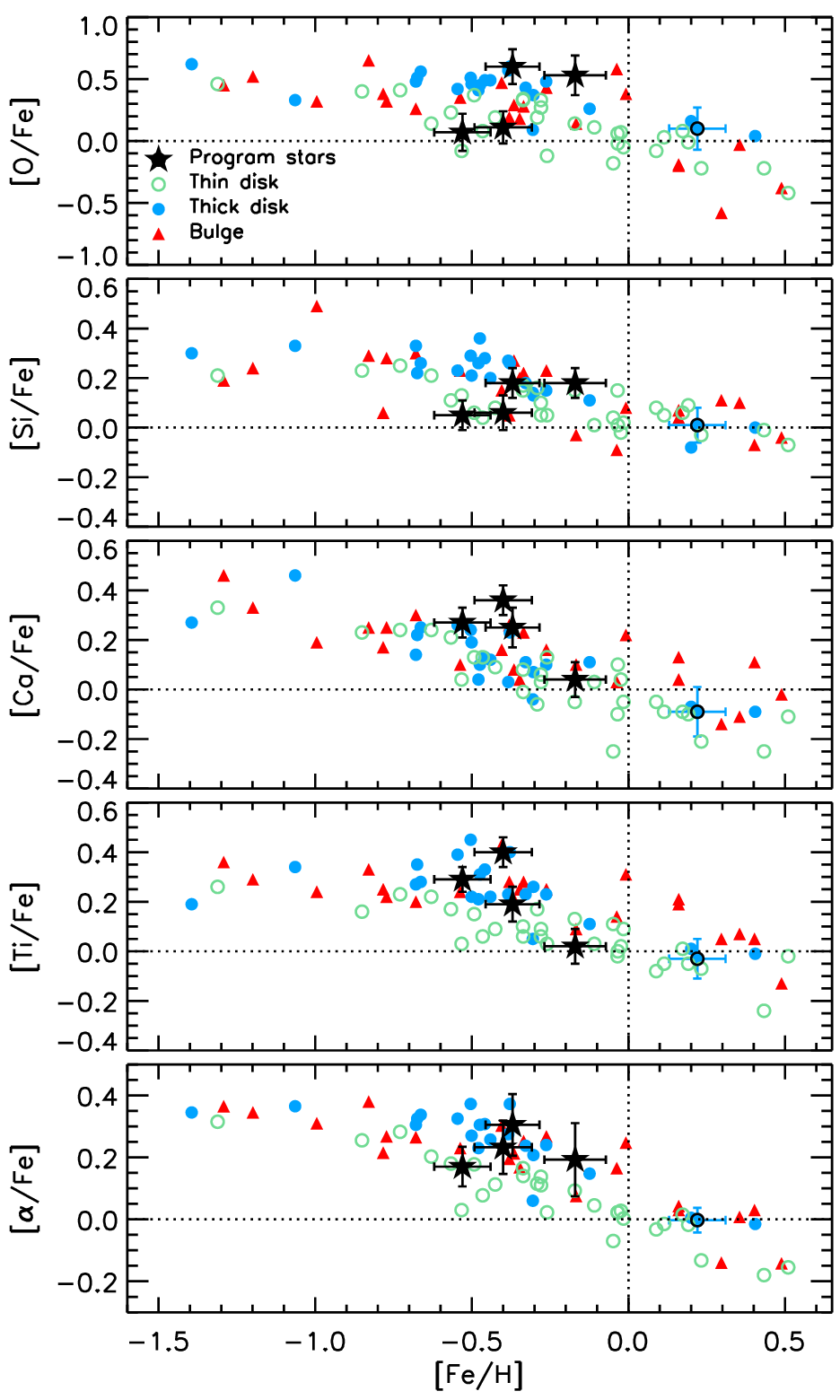} 
      \caption{Abundance [X/Fe] versus [Fe/H] for the program stars. For the
bottom panel, $\alpha$ is the average of O, Si, Ca and Ti. Thin disk (open aqua
circles), thick disk (filled blue circles) and bulge (red triangles) red giant
stars from \citet{Alves-Brito:2010aa} are overplotted in each panel. The thick
disk comparison star HD 40409 is located at [Fe/H] = +0.20 and includes an
error bar. 
      \label{fig:alpha} }
\end{figure}

Examination of the individual $\alpha$-elements (O, Si, Ca and Ti), however,
reveals subtle differences among those elements (see Figure \ref{fig:alpha}).
For O and Si, the two most metal-rich stars have higher [X/Fe] ratios when
compared to the two most metal-poor stars. For Ca and Ti, however, the
situation is reversed in that the two most metal-rich stars have lower [X/Fe]
ratios. And therefore on average, all stars have similar [$\alpha$/Fe] ratios. 

Figure \ref{fig:alpha} enables us to compare the abundance ratios of the
program stars with the thin disk, thick disk and bulge red giant stars from
\citet{Alves-Brito:2010aa}\footnote{The main conclusion of that work was that
the bulge and local thick disk stars are chemically similar and that they are
distinct from the local thin disk.}. Recall that our analysis includes HD 40409
also studied by \citeauthor{Alves-Brito:2010aa} and that we have adjusted our
abundance scale to match theirs (at least for Fe and the $\alpha$-elements).
Therefore, we are confident that there are no major systematic abundance
offsets between the program stars and the comparison sample. The program stars
occupy the same region of chemical abundance space as the comparison thin disk,
thick disk and bulge objects. It is not obvious, however, whether the program
stars more closely follow the thin disk or thick disk abundance trends due to
the different behaviour of the four $\alpha$-elements. Nevertheless, based on
these $\alpha$-elements, we conclude that there are no unusual chemical
abundance patterns amongst the program stars. 

Our analysis also included the Fe-peak elements Cr, Ni and Cu. For these
elements, the program stars lie on, or very near, the well-defined trends
exhibited by local red giant stars of comparable metallicity (e.g.,
\citealt{Luck:2007aa}).  Unlike the $\alpha$-elements, however, the possibility
exists that there may be systematic abundance offsets between our values and
the literature comparison sample for these three elements as we have no stars
in common. 

We measured abundances for the neutron-capture elements Ba, La and Eu. The
first two elements are produced primarily through the $s$-process while the
latter is an $r$-process element. Observations of open cluster giants indicate
that the Ba abundance increases with decreasing age \citep{DOrazi:2009aa}. Such
a chemical signature was interpreted as being due to extra contributions from
low-mass stars to the Galactic chemical evolution. Abundance trends with age,
however, are not seen for other $s$-process elements such as Zr and La
\citep{Jacobson:2013aa}. We find [Ba/Fe]~$\lesssim$~0.15 and in the
context of the \citet{DOrazi:2009aa} results, the program stars resemble open
clusters with ages $>$ 2 Gyr. Within our limited sample, however, the [Ba/Fe]
ratio appears to {\it decrease} with increasing mass. Assuming mass is a proxy
for age, then this trend between abundance and age would be opposite to that
seen among the open clusters. More data are needed to examine this intriguing
result. The program stars all have enhanced [Eu/Fe] ratios, and high ratios
would be expected given that Eu and the $\alpha$ elements typically follow each
other (e.g., \citealt{Woolf:1995aa,Sakari:2011aa}). That said, the [Eu/Fe]
ratios are slightly higher than [$\alpha$/Fe], although the former are based on
a single line. 

\begin{figure}
\centering
      \includegraphics[width=.99\hsize]{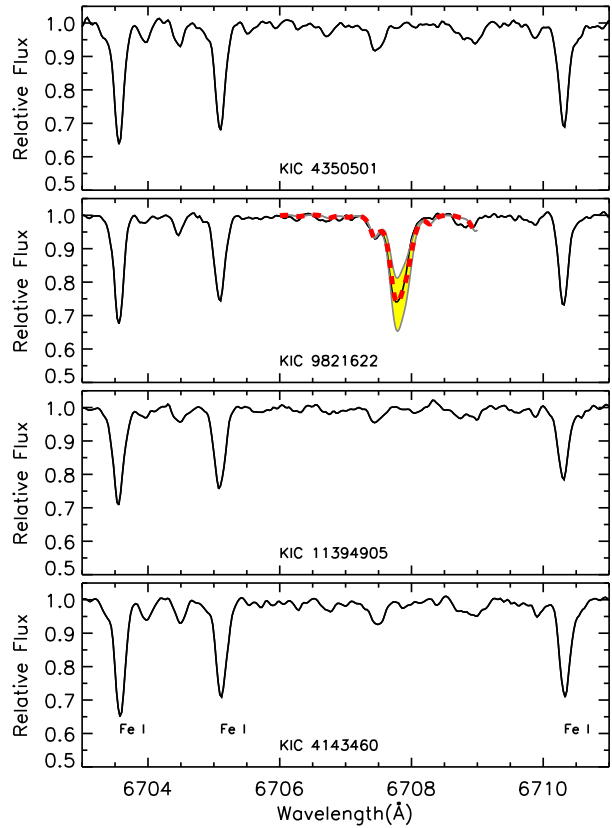} 
      \caption{Spectra near the 6707.8\,\AA\ Li line for the program stars. In
the second panel, we overplot the best fitting synthetic spectra (red dashed
lines) corresponding to A(Li)$_{\rm NLTE}$ = 1.76. The shaded yellow region
corresponds to synthetic spectra which differ from the best fit by $\pm$ 0.2
dex. The spectra were corrected for their heliocentric radial velocity and the
locations of some nearby \fei\ lines are indicated in the lower panel. 
\label{fig:specli} } 
\end{figure}

Finally, we measured lithium abundances (or limits) for the program stars using
spectrum synthesis (see Figure \ref{fig:specli}). Only KIC 9821622
(J19083615+4641212) has a detectable 6707\,\AA\ lithium line and we measure
A(Li)$_{\rm LTE}$ = 1.63 and A(Li)$_{\rm NLTE}$ = 1.76 using the non-LTE
corrections from \citet{Lind:2009aa}. While this star appears to be
lithium-rich when compared to the other program stars, the degree of enrichment
is considerably smaller than the highest values found in some giant stars,
A(Li)$_{\rm LTE}$ $\simeq$ 4 \citep{Reddy:2005aa}. For the other three stars,
the lithium abundance limits, A(Li)$_{\rm LTE}$ $\lesssim$ 0.4, overlap with
the limits in giant stars presented by \citet{Luck:2007aa}. 

 \citet{Jofre:2015aa} also studied KIC 9821622 using the GRACES spectra
available at the Gemini website (recall that our analysis is based on spectra
from an updated version of the OPERA data reduction pipeline). With the
exception of \teff, their stellar parameters \teff/\logg/\vt/[Fe/H] =
4725/2.73/1.12/$-$0.49 are in fair agreement with ours values
\teff/\logg/\vt/[Fe/H] = 4895/2.71/1.17/$-$0.40. For \teff, the difference is
170K; their values are derived from excitation equilibrium of \fei\ lines while
we employed the infrared flux method. For the elements in common between the
two studies (Li, O, Si, Ca, Ti, Cr, Fe, Ni, Ba La and Eu), the abundance ratios
are in good agreement with a mean difference of 0.00 $\pm$ 0.05 dex
($\sigma$ = 0.17). All elements agree to within 0.20 dex between the two
studies with the exception of O and Si for which the differences in [X/Fe] are
0.38 and 0.24 dex, respectively. Stellar parameters (\teff), line
selection and/or atomic data are the likely causes of the abundance
differences.

\section{DISCUSSION} 

\subsection{Chemical abundances}

All chemical abundance ratios appear ``normal'' when compared to local red giant
stars of similar metallicity. That is, our program stars exhibit no unusual
chemical abundance signatures that could provide clues to the origin of these
unusually massive and young stars with enhanced [$\alpha$/Fe] ratios. (We
will return to the lithium-rich object later in the discussion.) Given the
chemical similarities between local thick disk stars and those of the inner
disk and bulge \citep{Alves-Brito:2010aa,Bensby:2010aa}, it is difficult to use
chemical abundances to test the scenario proposed by \citet{Chiappini:2015aa}
in which these young [$\alpha$/Fe]-rich objects were formed near the Galactic
bar and migrated to their current locations. 

\subsection{Line broadening} 

Line broadening offers another possible clue to the origin of the program
stars. In particular, high line broadening could arise as the result of mass
transfer and/or stellar mergers and these processes are relevant in the
context of the blue straggler explanation proposed by \citet{Martig:2015aa}. We
note that the APOGEE pipeline does not measure broadening. We measured the line
broadening from spectrum synthesis and $\chi^2$ minimisation for five lines
(\cui\ 5105.50\,\AA, \cui\ 5782.14\,\AA, \baii\ 5853.69\,\AA, \baii\
6141.73\,\AA, \baii\ 6496.91\,\AA). These lines were already used in our
abundance analysis and are not too weak, nor too strong, such that reliable
measurements should be obtained. The broadening values\footnote{These
values are the FWHM of a Gaussian function applied to the synthetic spectra and
the line list includes isotopic shifts, hyperfine structure and blends.}
spanned a narrow range from 7.0 $\pm$ 0.4~\kms\ (KIC 4350501) to 7.7 $\pm$
0.2~\kms (KIC 11394905).  For a given star, the broadening values from the five
atomic lines were in very good agreement. We emphasise that our measurements
were obtained by fitting a single Gaussian function which incorporates the
($i$) instrumental profile, ($ii$) rotation and ($iii$) macroturbulent
velocity. For the GRACES spectra, the instrumental broadening alone accounts
for 4.4~\kms\ and thus the combined macroturbulent velocity and rotational
broadening ranges from 5.5 to 6.3~\kms, and such values appear normal for giant
stars \citep{Carney:2003aa,Hekker:2007aa,Luck:2007aa}. Therefore, none of the
program stars appear to be unusually broad lined. 

\subsection{Radial velocities and kinematics} 

As noted by \citet{Martig:2015aa}, these young [$\alpha$/Fe]-rich stars do not
possess unusual kinematic properties when compared to the other
[$\alpha$/Fe]-rich objects. We measured heliocentric radial velocities from the
observed wavelengths of the lines used in the EW analysis (see Table
\ref{tab:rv}). For three stars (KIC 4350501, KIC 9821622 and
KIC 4143460), our measured radial velocities are in excellent agreement
with the APOGEE values, i.e., these stars exhibit no evidence for radial
velocity variation beyond $\sim$1 \kms. 

\begin{table}
 \centering
 \begin{minipage}{190mm}
  \caption{Heliocentric radial velocities (\kms).} 
  \label{tab:rv} 
  \begin{tabular}{@{}lcrcr@{}}
  \hline
        Name & 
        Date & 
        RV & 
        $\sigma$RV & 
	APOGEE \\  
\hline 
  \hline
KIC 4350501  &  4 Jun 2015 & $-$83.4 & 0.5  & $-$83.3  \\
KIC 9821622  & 21 Jul 2015 &  $-$6.0 & 0.4  &  $-$5.5  \\
KIC 11394905 & 21 Jul 2015 & $-$69.8 & 0.5  & $-$75.5  \\
KIC 4143460  & 21 Jul 2015 &    +6.6 & 0.5  &    +6.6  \\
  \hline
\end{tabular}
\end{minipage}
\end{table}

For one star, KIC 11394905, there is evidence for a $\sim$6~\kms\ radial
velocity variation between the APOGEE and GRACES spectra, suggesting the
presence of at least one binary in our sample. In the context of the blue
straggler origin, the binary fraction offers a key diagnostic. Based on radial
velocity monitoring of the M67 open cluster, the blue straggler binary
frequency is 79 $\pm$ 24\%  which is significantly higher than the binary
frequency of 22.7 $\pm$ 2.1\% for the remaining M67 objects
\citep{Geller:2015aa}. Radial velocity measurements of metal-poor field blue
stragglers also reveal a high binary fraction of 47 $\pm$ 10\%
\citep{Carney:2001aa}; the same group find a binary fraction of $\sim$16\%
among ``normal'' metal-poor stars \citep{Carney:2003aa}. 

\citet{Martig:2015aa} dismissed the blue straggler origin on the following
grounds. They estimated the number of evolved blue stragglers to be a
factor of 3-4 lower than the young [$\alpha$/Fe]-rich stars in their sample.
That said, there are selection biases for the $Kepler$ sample as well as for
the subset observed by APOGEE that need to be taken into account. Additionally,
\citet{Martig:2015aa} found no evidence for anomalous surface rotation which
some blue stragglers possess. Finally, they noted that the radial velocity
variation among the APOGEE spectra for individual stars was small, $\sigma$RV
$<$ 0.2 \kms. 

For the four program stars, however, only two (KIC 4350501 and
KIC 4143460) have multiple radial velocity measurements from APOGEE; both
have three measurements with a baseline of $\sim$30 days. Given that
metal-poor blue stragglers tend to have long periods, $>$ 100 days, and
semi-amplitudes of $\sim$10 \kms\ \citep{Carney:2001aa}, we suggest that it is
unlikely that APOGEE would have detected radial velocity variations, if
present, over such a short baseline. 

\begin{figure}
\centering
      \includegraphics[width=.99\hsize]{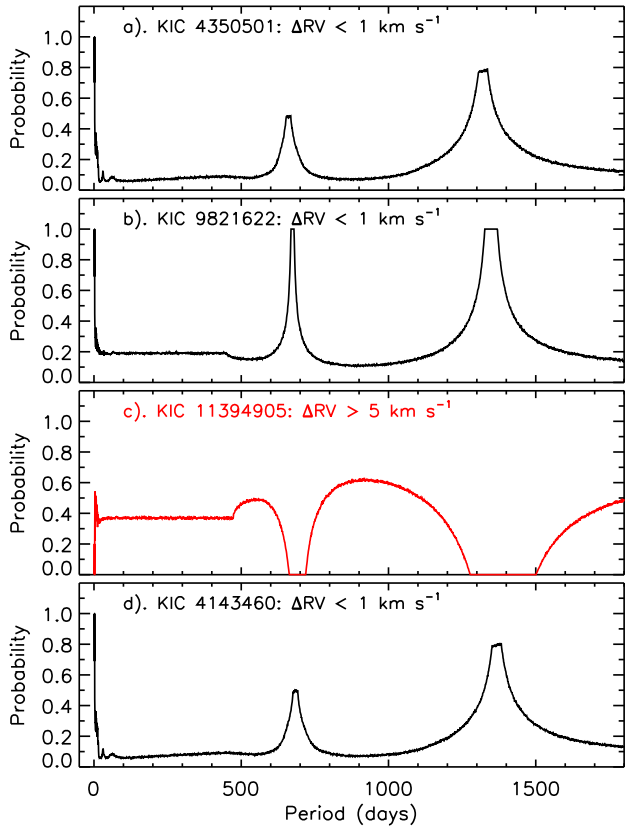} 
      \caption{Probability of detecting radial velocity amplitudes $\le$ 1
\kms\ (panels a, b and d) or $\ge$ 5 \kms\ (panel c) as a function of period
for the program stars. These values were based on Monte Carlo simulations
assuming a semi-amplitude of 10 \kms, circular orbits, the observed number of
epochs and their time spans (see text for details). The peaks in probability
correspond to the differences in dates when the stars were observed by APOGEE
and GRACES. 
      \label{fig:monterv} }
\end{figure}

Combining the APOGEE radial velocities with those measured from the GRACES
spectra, we now have additional epochs and a longer baseline over which to
examine the likelihood of detecting radial velocity variation. Following
\citet{Norris:2013aa}, we can then ask the following question: What is the
probability of observing a radial velocity variation $\le$ 1.0\footnote{For
three of the four program stars, the differences in radial velocities
between APOGEE and GRACES are below 1 \kms. While the uncertainties in the
APOGEE and GRACES radial velocities are $\sim$ 0.5 \kms, we do not know for
certain whether the zero-points are the same. We therefore conservatively adopt
a threshold velocity difference of 1 \kms\ when exploring the current
observational constraints on binarity.} \kms\ given the observed number of
epochs and their time spans? Using Monte Carlo simulations, we estimated these
probabilities in the following way. We assumed that each star had a circular
orbit with a semi-amplitude of 10 \kms; such values appear typical for blue
stragglers \citep{Carney:2001aa}. We tested all periods from 0.5 days to 30
days (in steps of 0.5 days) and then from 30 days to 1800 days (in steps of 1
day). For a given assumed period, we performed 10,000 realisations in which the
inclination angle was randomly set and the first ``observation'' was set at a
random phase. For all epochs of observation, we could obtain velocities. We
then asked the question: For what fraction of realisations is the maximum
velocity difference $\le$ 1.0 \kms? We plot those results in panels a, b and d
in Figure \ref{fig:monterv}. Given the small number of radial velocity
measurements, we cannot exclude the possibility that these stars are binaries.
The high probability peaks near 700 and 1350 days are caused by the typical
baseline between the first APOGEE measurements and the GRACES observations. 

For KIC 11394905 (panel c in Figure \ref{fig:monterv}), the two radial
velocity measurements differ by $\sim$6 \kms\ and we asked a slightly
different question: For what fraction of realisations is the maximum velocity
difference $\ge$ 5.0\footnote{Again, we do not know whether the APOGEE and
GRACES measurements have the same zero points, so we conservatively adopt a
threshold value of 5 \kms\ in this exercise.} \kms? The limited observations
can only preclude periods around 700 and 1400 days. 

Informed by the above simulations, we conclude that all program stars could be
evolved blue stragglers and we do not have sufficient radial velocity
measurements to exclude binarity.  Therefore, long-term radial velocity
monitoring for the entire sample from \citet{Martig:2015aa} is essential to
establish any radial velocity variation and thereby place stronger constraints
on the blue straggler hypothesis. 

\citet{Preston:2000aa} found no evidence for $s$-process enhancements among
their sample of long period low eccentricity blue stragglers. Similarly, our
sample also exhibit no evidence for $s$-process element enrichment. 

Finally, recall that one star, KIC 9821622, appears to be lithium rich. The
exact process that causes enhanced Li abundances in a small fraction of evolved
stars has not been identified (e.g., \citealt{Charbonnel:2000aa}), and the
heterogeneity in evolutionary phase among Li-rich giants suggests that multiple
mechanisms may be at work \citep{Martell:2013aa}. If our sample are evolved
blue stragglers, then the mechanism(s) responsible for lithium enrichment must
also operate in these objects. 

\subsection{Spectral energy distributions}

\begin{figure*}
\centering
\includegraphics[width=.85\hsize]{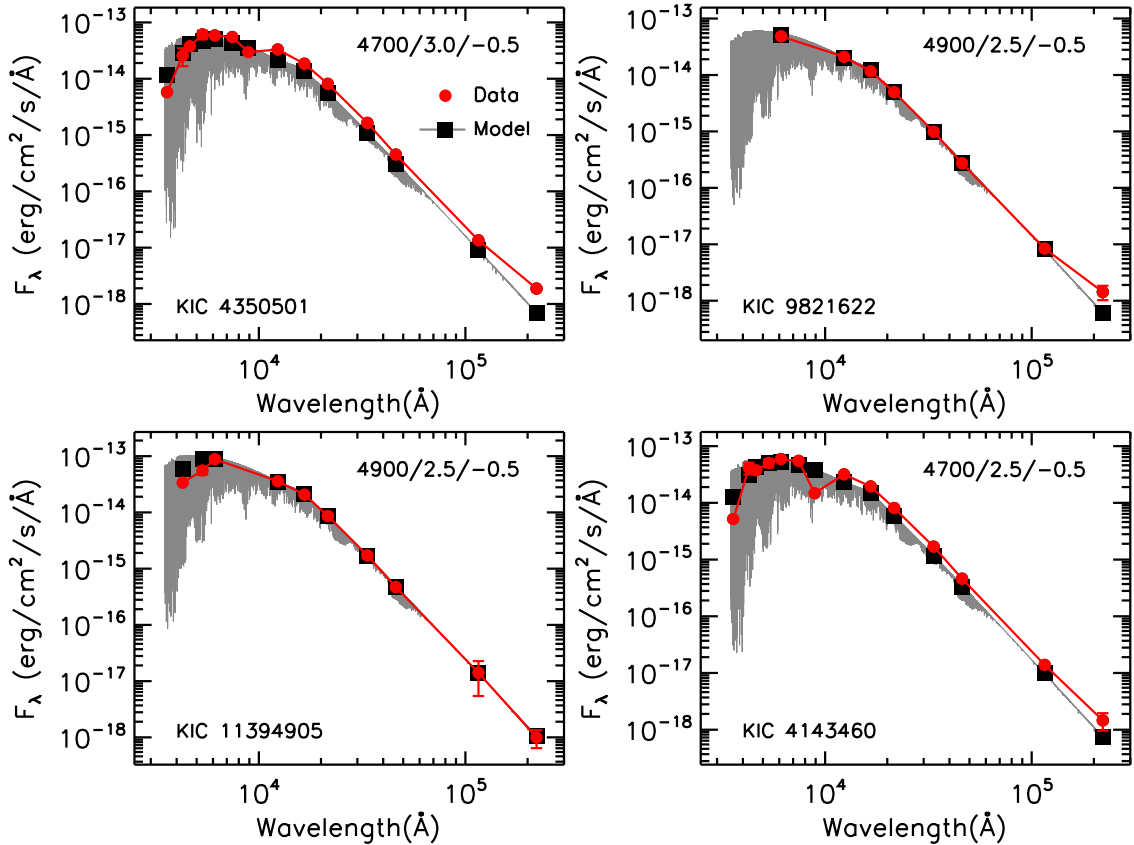} 
\caption{Spectral energy distributions for the four program stars (red
circles). The best fitting models (black squares) and theoretical spectra (grey
lines) are overplotted. The model parameters (\teff/\logg/[Fe/H]) are indicated
in each panel. (See text for details on the SEDs and the fitting.) 
\label{fig:sed} }
\end{figure*}

To further explore the possibility that these stars could be blue stragglers,
or binaries in general, we examine their spectral energy distributions. To
reiterate, the key aspect we are focusing upon is the possibility that the
program stars are binaries such that the inferred masses are high due to mass
transfer or merger leading to ages that are underestimated for a single star.
Spectral energy distributions (SEDs) were created using the Spanish Virtual
Observatory SED Analyzer (VOSA)\footnote{{\url
http://svo2.cab.inta-csic.es/theory/vosa/}} \citep{Bayo:2008aa}. For the
program stars, the SEDs are generated using photometry from SDSS DR9
\citep{Ahn:2012aa}, Tycho-2 \citep{Hog:2000aa}, 2MASS \citep{Majewski:2003aa}
and WISE \citep{Wright:2010aa} and fit using the BT-NextGen (AGSS2009) grid of
stellar model atmospheres created by \citet{Allard:2012aa}. (For 4350501, the
22 $\mu$m WISE W4 bandpass was not included in the fit.) In Figure
\ref{fig:sed}, we find that two stars (KIC 9821622 and KIC 4350501) exhibit an
infrared (IR) excess and a third star (KIC 4143460) may also show an IR excess.
The IR excess is most notable in the 22 $\mu$m WISE W4 bandpass. 

Among red giant stars, $<$1\% exhibit an IR excess
\citep{Jones:2008aa,Bharat-Kumar:2015aa}. On the other hand, IR excesses are
commonly found in post-AGB, RV Tauri, and Lambda Bootis stars (e.g.,
\citealt{Van-Winckel:1995aa,Giridhar:2005aa}). These objects are (likely)
binary systems with debris disks or dusty circumstellar environments that have
undergone dust-gas winnowing (see \citealt{Venn:2014aa}). An examination of all
14 stars in the \citet{Martig:2015aa} sample show that only 5 stars have clear
IR excesses, including 3 of these stars with GRACES spectra. 

\section{CONCLUSIONS} 

We have analysed high-resolution spectra of four massive (i.e., young)
[$\alpha$/Fe]-rich stars from \citet{Martig:2015aa} obtained using
Gemini-GRACES during the 2015 on-sky tests. While one object appears to be
lithium rich, we find no chemical abundance anomalies among the program stars
when compared to local giants. Although only one of the four stars exhibits a
radial velocity variation, given the small number of radial velocity
measurements, we cannot exclude the possibility that the three remaining stars
are binaries. \citet{Martig:2015aa} suggested that these young
[$\alpha$/Fe]-rich stars could be evolved blue stragglers. The spectral energy
distributions indicate that two (and perhaps three) of the four stars exhibit
an infrared excess, characteristic of certain types of binary stars. In light
of the high $>$ 50\% binary fraction among blue stragglers, long-term radial
velocity monitoring is essential to test this scenario. 

\section*{Acknowledgments}

We thank the anonymous referee for helpful comments.
D.Y thanks John E.\ Norris for helpful discussions. 
D.Y, L.C, M.A, A.D and K.S gratefully acknowledge support from the Australian
Research Council (grants FL110100012, DP120100991, FT140100554 and
DP150100250). 
S.M. gratefully acknowledges support from the Australian Research Council
(grant DE 140100598). 
JM thanks support by FAPESP (2010/50930-6).

\label{lastpage}

\end{document}